# Potassium-doped BaFe$_2$As$_2$ superconducting thin films with a transition temperature of 40 K

Nam Hoon Lee,[a)] Soon-Gil Jung,[a)] Dong Ho Kim,[b)] and Won Nam Kang[a,*]

[a)]BK21 Physics Division and Department of Physics, Sungkyunkwan University, Suwon 440-746, Republic of Korea

[b)]Department of Physics, Yeungnam University, Gyeongsan 712-749, Republic of Korea

Abstract

We report the growth of potassium-doped BaFe$_2$As$_2$ thin films, where the major charge carriers are holes, on Al$_2$O$_3$ (0001) and LaAlO$_3$ (001) substrates by using an *ex-situ* pulsed laser deposition technique. The measured $T_c$'s are 40 and 39 K for the films grown on Al$_2$O$_3$ and LaAlO$_3$, respectively and diamagnetism indicates that the films have good bulk superconducting properties below 36 and 30 K, respectively. The X-ray diffraction patterns for both films indicated a preferred *c*-axis orientation, regardless of the substrate structures of LaAlO$_3$ and Al$_2$O$_3$. The upper critical field at zero temperature was estimated to be about 155 T.

Since the discovery of iron-based superconductors having a superconducting transition temperature ($T_c$) of 26 K in 2008, many groups have been interested in these kinds of materials, and a tremendous number of iron-based superconductors have been discovered so far.[1-10] Generally, iron-based superconductors can be classified as *RE*FeAsO (*RE*: rare earths), *AE*Fe$_2$As$_2$ (*AE*: alkaline earths), *A*FeAs (*A*: Li, Na), and FeCh (Ch: chalcogens). Various researchers have tried to fabricate iron-pnictide superconducting thin films,[11-18] but only a few superconducting films of iron-based superconductors have been successfully fabricated until now. For a comparative study on superconductivity between the electron-doping and the hole-doping cases based on films, the fabrication of hole-doped pnictide superconducting films is crucial. All however, iron-based superconducting film reported so far have been electron-doped compounds, such as Te-doped FeSe,[12] Co-doped *AE*Fe$_2$As$_2$ (*AE*: Sr, Ba),[13-16] and F-doped *RE*FeAsO (*RE*: La, Sm),[17] and the highest $T_c$ reported in Fe-based films[13] has been 24.5 K in Co-doped BaFe$_2$As$_2$, which is much lower than that (39 K) of hole-doped BaFe$_2$As$_2$ compounds. Thus, hole-doped superconducting films are required to understand the high-$T_c$ superconductivity of iron-pnictide superconductors. Growth of hole-doped superconducting thin films is known to be very difficult because of the high volatility of the doping elements, such as potassium and sodium. The huge difference of the vapor pressures in the elements contained in the materials causes further difficulty when fabricating films.[7,8,19] The differences in the volatilities of compounds has been an issue[19-21] in the thin-film processing field for a long period, and the situation is the same for iron-based superconductors.

We have fabricated potassium-doped Ba$_{0.6}$K$_{0.4}$Fe$_2$As$_2$ thin film, in which the major charge carriers were holes, by using both a pulsed laser deposition (PLD) technique with homemade targets having stoichiometric ratio and a postannealing process. The films deposited on $Al_2O_3$ (0001) and $LaAlO_3$ (001) substrates showed superconducting critical temperatures ($T_c$) of 39 and 40 K, respectively, which are the highest critical temperatures reported in Fe-based superconducting films and which are as high as those of bulk samples.[7,22] The X-ray diffraction patterns for both films indicated a preferred *c*-axis orientation, regardless of the substrate structures of $LaAlO_3$ and $Al_2O_3$, which have cubic and hexagonal structures, respectively. These results suggest that our fabricating method should be a very effective way to grow epitaxial films.

The method of synthesizing targets is similar to that in our previous work.[16] The compositional molar ratio of Ba lumps (99%), Fe powder (99.9%), and As pieces (99.99%) were 0.6, 2, and 2.6, respectively. The mixtures, which were put into an alumina crucible, were sealed in quartz ampoule. It was annealed at 900 °C for 12 hours. After annealing, the reacted samples were ground and annealed in sealed quartz ampoules again at 900 °C for 12 hours. After the second annealing, the compounds were homogenized and pressed as a shape of a disk with a diameter of 15 mm. The pressed target was annealed at 800 °C for 8 hours after having been encapsulated in a quartz ampoule. The synthesized targets were ablated by using a pulsed laser at room temperature in a high vacuum of $10^{-6}$ Torr. The laser beam was generated by using a KrF excimer laser (Lambda Physik) with a 248 nm wavelength and the energy density of 5 J $cm^{-2}$ at a frequency of 48 Hz. The deposited precursor films without potassium were sealed in quartz ampoules with potassium lumps (99.5%) of about 0.6 g and annealed at 700 °C for 6 hours. The heating and the cooling rates were 100 °C $h^{-1}$ and 200 °C $h^{-1}$, respectively. Potassium was handled under a 99.999% argon atmosphere, and the quartz ampoules were sealed under high vacuum.

We used a $LaAlO_3$ (001) and $Al_2O_3$ (0001) substrates with dimensions of 10 mm ×10 mm × 0.5 mm for the growth of the potassium-doped $BaFe_2As_2$ superconducting films. The phases and the crystalline qualities of the potassium-doped $BaFe_2As_2$ films were investigated by using X-ray diffractometry. The thicknesses and the surface morphologies of the films were confirmed by using scanning electron microscopy. The superconducting properties were measured by using a physical property measurement system (PPMS, Quantum Design) and a vibrating sample measurement system (VSM, Quantum Design). The resistivity was measured by using a standard four-probe method.

The temperature dependences of the resistivity ($\rho$) for the potassium-doped $BaFe_2As_2$ films fabricated on $LaAlO_3$ (001) and $Al_2O_3$ (0001) substrates are shown in figure 1 (a). The inset of figure 1 (a) is an enlarged view near the superconducting transition temperature ($T_c$). The onset-transition temperatures of the films prepared on $LaAlO_3$ and $Al_2O_3$ were about 39 K and 40 K, with sharp superconducting transition widths ($\Delta T_c = T_{c,90\%} - T_{c,10\%}$) of 1.6 K and 1.7 K, respectively. The $T_c$ of the film grown on $Al_2O_3$ is higher than that of bulk samples, and that of the sample grown on $LaAlO_3$ is the same as that of the bulk samples.[7] The $T_c$ of 40 K is the highest critical temperature observed in films of Fe-based pnictide superconductors. The residual resistivities at 41 K for the films growth on $LaAlO_3$ and $Al_2O_3$ were 0.54 mΩ cm and 0.49 mΩ cm, with residual resistivity ratios (RRR = $\rho_{300K}/\rho_{41K}$) of 5.2 and 5.0, respectively. Their RRR values are comparable to that observed in bulk samples, but the residual resistivity is much higher than that in bulk samples.[7,22]

Figure 1 (b) shows the temperature dependences of the magnetization of potassium-doped BaFe$_2$As$_2$ films grown on Al$_2$O$_3$ and LaAlO$_3$ substrates which were measured at H = 10 Oe perpendicular to the film. The data are normalized to the absolute values at 5 K. The irreversible temperature of the film fabricated on Al$_2$O$_3$ was 37.5 K, which is higher than that of the film grown on LaAlO$_3$ and is consistent with the zero-transition temperature in the $\rho$-$T$ data. The clear diamagnetic signal and the clear separation of the field-cooled (FC) and the zero-field-cooled (ZFC) data in both samples indicate that they have good bulk superconducting properties. The small diamagnetic signal in the FC data reflects strong pinning nature of the film, which could originate from various defects or grain boundaries. According to the scanning electron microscopy (SEM) images of the surface shown in the inset of figure 1 (b), the potassium-doped BaFe$_2$As$_2$ films grown on LaAlO$_3$ substrates showed good grain connectivity, but further study is required to obtain high-quality films with smooth surface morphologies. The thicknesses of the films were 1 μm, which was obtained from the cross-sectional SEM images.

Figure 2 shows $\theta$-$2\theta$ scan X-ray diffraction spectra of the potassium-doped BaFe$_2$As$_2$ films deposited on LaAlO$_3$ (001) and Al$_2$O$_3$ (0001) substrates. Prominent (00$l$) peaks were detected in both films in spite of the different crystal structures of the Al$_2$O$_3$ (0001) and the LaAlO$_3$ (001) substrates. The preferred $c$-axis-oriented films, irrespective of the substrate, suggest that our *ex-situ* growth method would be a powerful technique for the fabrication of potassium-doped BaFe$_2$As$_2$ epitaxial films and that this method could be applied to grow other hole-doped iron-pnictide superconducting films.

We calculated the lattice parameters of films by using a least squares refinement method. The crystallographic data for the film grown on an Al$_2$O$_3$ substrate showed $a$ = 3.9177 Å and $c$ = 13.3313 Å, a $c/a$ ratio of 3.4028. The data for the film prepared on LaAlO$_3$ showed $a$ = 3.9068 Å, $c$ = 13.4037 Å, and $c/a$ = 3.4308. It is significant to compare these two types of films because these films showed different values of $T_c$ and the film on Al$_2$O$_3$ had a critical temperature that was 1 K higher than that (39 K) of bulk samples. The $c$-axis lattice constant of the film on LaAlO$_3$ was larger than that of the film on Al$_2$O$_3$ whereas the $a$-axis of the film on Al$_2$O$_3$ was larger than that of the film on LaAlO$_3$, suggesting that the film on Al$_2$O$_3$ is strained in the $a$-axis direction compared with the film on LaAlO$_3$. It has been reported that the value of $T_c$ could be increased with increasing strain in the $a$-axis in Co-doped BaFe$_2$As$_2$.[14] This is one of the possible interpretation for the film on Al$_2$O$_3$ having a higher $T_c$ than the film on LaAlO$_3$.

The temperature dependences of the resistivity of a potassium-doped BaFe$_2$As$_2$ film fabricated on an Al$_2$O$_3$ substrate for various values of the magnetic fields from 0 to 7 T are shown in Figure 3. The inset presents the upper critical field as a function of temperature [$H_{c2}(T)$] near the transition temperature. The $H_{c2}$ values were defined as the field at which the resistivity dropped to 90% ($\rho_{90\%}$), 50% ($\rho_{50\%}$), and 10% ($\rho_{10\%}$) of its normal-state value at 41 K. The d$H_{c2}$/d$T$ values evaluated from the 90, 50, and 10% drop offs from the normal-state resistivity were −5.67 T K$^{-1}$, −3.84 T K$^{-1}$, and −3.22 T K$^{-1}$, respectively. The $H_{c2}$ at zero temperature which were estimated using the Werthamer–Helfand–Hohenberg formula, $H_{c2}(0)$ = −0.69 $T_c$(d$H_{c2}$/d$T$)$_{T=T_c}$, was ~ 155 T for the $\rho_{90\%}$ criterion, which is about 20% smaller that that of a single-crystalline sample.[22]

In summary, we report the growth of potassium-doped BaFe$_2$As$_2$ superconducting films on Al$_2$O$_3$

(0001) and LaAlO$_3$ (001) substrates by using an *ex-situ* pulsed laser deposition technique with a KrF excimer laser. The $T_c$ of 40 K for our samples is the highest one among those reported for iron-based superconducting films. The irrelevance of the substrate structure and the growth orientation direction indicates the superiority of this method for fabricating high-quality films. The potassium-doped BaFe$_2$As$_2$ films grown by using this *ex-situ* process showed good bulk superconducting properties, and the upper critical field at zero temperature were estimated to be as high as 155 T.

## ACKNOWLEDGMENTS

This work was supported by the Korea Science and Engineering Foundation (KOSEF) grant funded by the Korea government (MEST, No. R01-2008-000-20586-0).

## REFERENCES

[1] Y. Kamihara, T. Watanabe, M. Hirano, and H. Hosono, J. Am. Chem. Soc **130**, 3296 (2008).

[2] H. Takahashi, K. Igawa, K. Arii, Y. Kamihara, M. Hirano, and H. Hosono, Nature (London) **453**, 376 (2008).

[3] H-H. Wen, G. Mu, L. Fang, H. Yang, and X. Zhu, Europhys. Lett. **82**, 17009 (2008).

[4] Z. A. Ren, J. Yang, W. Lu, W. Yi, X-L. Shen, Z–C. Li, G–C. Che, X–L. Dong, L–L. Sun, F. Zhou, and Z–X. Zhao, Europhys. Lett. **82**, 57002 (2008).

[5] X. H. Chen, T. Wu, G. Wu, R. H. Liu, H. Chen, and D. F. Fang, Nature (London) **453**, 761 (2008).

[6] J. Yang, Z–C. Li, W. Lu, W. Yi, X–L. Shen, Z–A. Ren, G–C. Che, X–L. Dong, L–L. Sun, F. Zhou, and Z.–X. Zhao, Supercond. Sci. Technol. **21**, 082001 (2008).

[7] M. Rotter, M. Tegel, and D. Johrendt, Phys. Rev. Lett. **101**, 107006 (2008).

[8] G–F. Chen, Z. Li, G. Li, W–Z. Hu, J. Dong, J. Zhou, X–D. Zhang, P. Zheng, N–L. Wang, and J–L. Luo, Chinese Phys. Lett. **25**, 3403 (2008).

[9] A. Leithe-Jasper, W. Schnelle, C. Geibel, and H. Rosner, Phys. Rev. Lett. **101**, 207004 (2008).

[10] A. S. Sefat, R. Jin, M. A. McGuire, B. C. Sales, D. J. Singh, and D. Mandrus, Phys. Rev. Lett. **101**, 117004 (2008).

[11] M. J. Wang, J. Y. Luo, T. W. Huang, H. H. Chang, T. K. Chen, F. C. Hsu, C. T. Wu, P. M. Wu, A. M. Chang, and M. K. Wu, Phys. Rev. Lett. **104**, 117002 (2009).

[12] Y. Han, W. Y. Li, L. X. Cao, S. Zhang, B. Xu, and B. R. Zhao, J. Phys.: Condens. Matter **21**, 235702 (2009).

[13] S. Lee, J. Jiang, Y. Zhang, C. W. Bark, J. D. Weiss, C. Tarantini, C. T. Nelson, H. W. Jang, C. M. Folkman, S. H. Baek, A. Polyanskii, D. Abraimov, A. Yamamoto, J. W. Park, X. Q. Pan, E. E. Hellstrom, D. C. Larbalestier, and C. B. Eom, Nature Mater., Published online: 28 Feb. doi:10.1038/nmat2721 (2010).

[14] K. Iida, J. Hänisch, R. Hühne, F. Kurth, M. Kidszun, S. Haindl, J. Werner, L. Schultz, and B. Holzapfel1, Appl. Phys. Lett. **95**, 192501 (2009).

[15] H. Hiramatsu, T. Katase, T. Kamiya, M. Hirano, and H. Hosono, Appl. Phys. Express **1**, 101702 (2008).

[16] E-M. Choi, S-G. Jung, N. H. Lee, Y-S. Kwon, W. N. Kang, D. H. Kim, M-H. Jung, S-I. Lee, and L. Sun, Appl. Phys. Lett. **95**, 062507 (2009).

[17] S. Haindl, M. Kidszun, A. Kauffmann, K. Nenkov, N. Kozlova, J. Freudenberger, T. Thersleff, J. Hänisch, J. Werner, E. Reich, L. Schultz, and B. Holzapfel, Phys. Rev. Lett. **104**, 077001 (2010).

[18] T. Katase, H. Hiramatsu, H. Yanagi, T. Kamiya, M. Hirano, and H. Hosono, Solid State Commun. **149**, 2121 (2009).

[19] S-G. Jung, N. H. Lee, W. K. Seong, W. N. Kang, E. -M. Choi, and S. -I. Lee, Supercond. Sci. Technol. **21**, 085017 (2008).

[20] T. Atsuki, N. Soyama, T. Yonezawa, and K. Ogi, Jpn. J. Appl. Phys. **34**, 5096 (1995).

[21] I–H. Kim, H–S. Park, Y–J. Park, and T. Kim, Appl. Phys. Lett. **73**, 1634 (1998).

[22] D. L. Sun, Y. Liu, and C. T. Lin, Phys. Rev. B **80**, 144515 (2009).

**FIGURES**

FIG. 1. (a) Resistivity-temperature and (b) normalized magnetization-temperature graphs of potassium-doped $BaFe_2As_2$ thin films. The normalized magnetization is measured at $H = 10$ Oe. The inset of figure (a) is a magnified view near the superconducting transition temperature. The film grown on $Al_2O_3$ (0001) shows a higher $T_c$ than the film deposited on $LaAlO_3$ (001). The clear diamagnetic signals in figure (b) indicate the good bulk superconductivity of our samples. The inset of figure 1 (b) shows the surface morphology of a film grown on a $LaAlO_3$ substrate.

FIG. 2. X-ray diffraction spectra ($\theta$-$2\theta$ scans) of potassium-doped $BaFe_2As_2$ thin films grown on $Al_2O_3$ (0001) and $LaAlO_3$ (001) substrates. These curves indicate that the films are preferentially oriented along the *c*-axis, regardless of the substrate.

FIG. 3. The temperature dependences of the resistivity of a potassium-doped $BaFe_2As_2$ thin film fabricated on an $Al_2O_3$ substrate, measured in fields up to 7 T. The inset show the temperature dependence of the upper critical magnetic field, where the upper critical magnetic field was determined as the field at where the resistivity dropped 90, 50, or 10% from its normal-state values at 41 K.

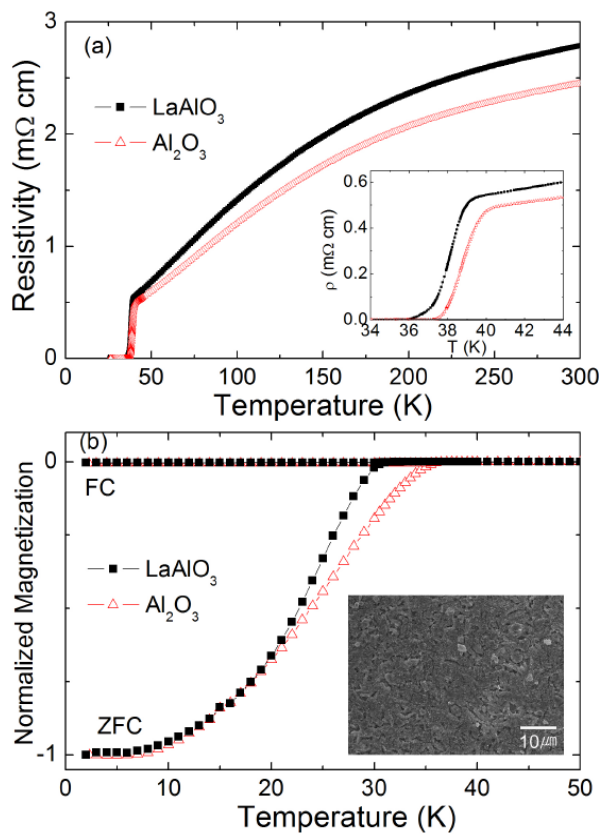

Fig. 1, N. H. Lee *et al.*, Potassium-doped BaFe$_2$As$_2$...

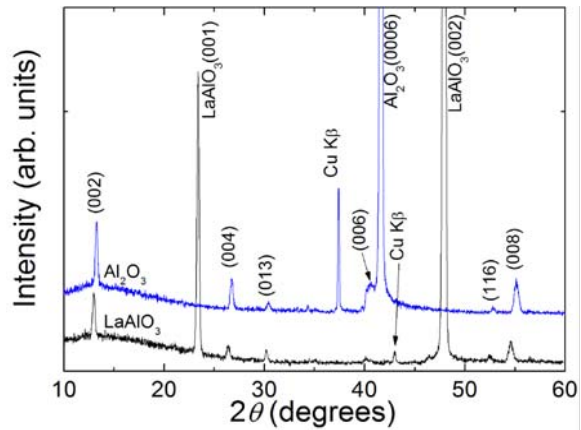

Fig. 2, N. H. Lee *et al.*, Potassium-doped BaFe$_2$As$_2$...

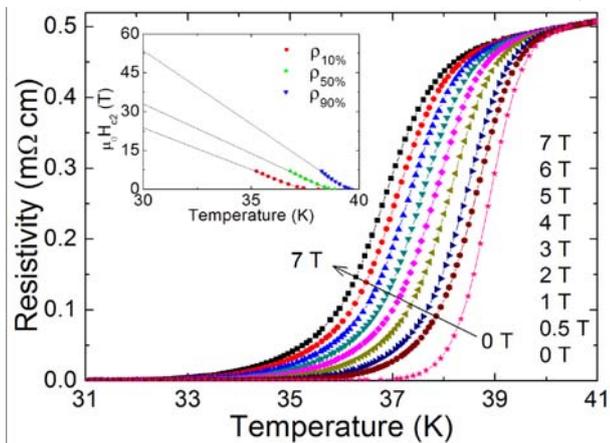

Fig. 3, N. H. Lee *et al.*, Potassium-doped BaFe$_2$As$_2$...